\documentclass[8pt,twocolumn,english]{extarticle}

\usepackage[T1]{fontenc}
\usepackage[latin9]{inputenc}
\usepackage{geometry}
\usepackage{prettyref}
\usepackage{float}
\usepackage{amsmath}
\usepackage{amsthm}
\usepackage{graphicx}
\usepackage{setspace}
\makeatletter

\providecommand{\tabularnewline}{\\}
\floatstyle{ruled}
\newfloat{algorithm}{tbp}{loa}
\providecommand{\algorithmname}{Algorithm}
\floatname{algorithm}{\protect\algorithmname}

\numberwithin{equation}{section}
\numberwithin{figure}{section}
\newenvironment{lyxcode}
	{\par\begin{list}{}{
		\setlength{\rightmargin}{\leftmargin}
		\setlength{\listparindent}{0pt}%
		\raggedright
		\setlength{\itemsep}{0pt}
		\setlength{\parsep}{0pt}
		\normalfont\ttfamily}%
	 \item[]}
	{\end{list}}

\usepackage{babel}
\usepackage{algorithm}
\usepackage{algpseudocode}
\usepackage{babel}
\usepackage{babel}

\makeatother

\usepackage{babel}
\usepackage[bibstyle=authoryear,citestyle=dtk,backend=bibtex8]{biblatex}
\begin{document}
\title{It is not just about the Melody: How Europe Votes for its Favorite
Songs}
\author{Anej Svete, anej.svete@protonmail.com\\
Jakob Hostnik, jh9934@student.uni-lj.si\\
University of Ljubljana, Faculty of Computer and Information Science,
\\
Ve\v{c}na pot 113, 1000 Ljubljana, Slovenia}
\maketitle
\begin{abstract}
The Eurovision Song Contest is a popular annual international song
competition organized by the European Broadcasting Union. The winner
is decided by the audience and expert juries from each participating
nation, which is why the analysis of its voting network offers a great
insight into what factors, besides the quality of the performances,
influence the voting decisions.

In this paper, we present the findings of the analysis of the voting
network, together with the results of a predictive model based on
the collected data. We touch upon the methodology used and describe
the dataset we carry the analysis on. The results include some general
features of the voting networks, the exposed communities of countries
that award significantly more points among themselves than would be
expected and some predictions on what the biggest factors that lead
to this phenomenon are. We also include the model to predict the votes
based on network structure of both previous votes and song preferences
of nations, which we found not to offer much improved predictions
than relying on the betting tables alone. 
\end{abstract}

\section{Introduction\label{sec:Introduction}}

The Eurovision Song Contest (ESC) has been held every year since 1956.
Its initial purpose was to unite the European nations after the Second
World War and has since evolved into an annual entertainment spectacle
followed by millions of people. Every rendition of the contest except
for the first featured one song entry by each participating country.
The countries involved are mostly European, with the recent addition
of some nations from outside the continent, e.g. Israel and Australia.

Although some rules have changed throughout the years, the main principles
of the competition remain the same. Each participating nation awards
some number of points to the performances chosen by the public and
jury of that country. Countries can not give points to themselves.
The song with the highest number of points wins. The current system
is in place since 2004 and consists of two semi-finals and a final.
Each country gets the same number of points to distribute, and they
are equally split between the jury and televoting votes. Both are
converted on a scale of points ranging from 1 to 12, with the exception
of 9 and 11 points, which are not awarded. This means that every country
awards points to 10 performances (\citetitle{ebu,esc_wiki}).

The nature of voting offers a intriguing opportunity to explore what
European countries base their voting decisions on. Some of the commonly
attributed factors include geographical proximity, language similarity
(\textcite{dekker}), ethnic structure (\textcite{SpierdijkGeography}),
common history, political preference and cultural similarity (\textcite{escpolitical}).
We try to extract as much information on the deciding factors as possible
from the network structure and the nation's attributes to pinpoint
the most important influences and to leverage that information to
infer future voting choices. As discussed in \prettyref{sec:Related-work},
we have not found any previous work that tried to use the network
information for future predictions. It it a challenging task, partly
because of the challenge of obtaining enough quality data and partly
because of the changing format of the competition. We take these historical
differences into account during the analysis. However, since one of
our main goals is making future predictions, the most recent results
are the most important, and these are enough to expose some main trends.

Besides making predictions, we are also interested in how different
similarities between nations correlate to their voting patterns. Just
by paying some attention to points distribution, it can be seen that
there is some correlation between the points awarded and geographic
proximity. We try to leverage the network features (e.g. the strength
of bias shown throughout the years and the community structure this
implies) to extract useful information more precisely and then reason
about the biggest deciding factors on the voting. To achieve that,
we perform community detection on the network of shown bias to infer
the influences.

There have also been some suggestions that building these friendships
allows the participants to achieve better scores and rank higher in
the competition. One part of the paper thus also focuses on finding
out if this is really the case by finding a correlation between the
community structure of specific nations and their success in the competition.

To prevent too much biased voting, some measures have already been
taken by the ESC committee. For example, they try to minimize the
number of neighboring countries competing in the same semi-final and
since only the countries performing that night can vote, neighboring
countries have less of a chance to help each other get into the final
(\citetitle{ebu,esc_wiki}). This measure can of course not be taken
in the final.

One thing that we also take a look into is the notion of neglect between
countries. By this we mean the behavior when countries which are somehow
linked to one another seldom award each other a significant number
of points. In other words, neighbors that do not exchange points could
be regarded as neglecting each other.

When the major influences on the voting behavior are exposed, we turn
our attention to the actual predictor of future voting behavior and
use the information about the communities in the second part of the
project together with some additional data that we learn through song
and artist features to try to infer the number of points countries
will award in future competitions. The model is used together with
betting predictions since they are considered to be the best existing
way to predict the outcome of the competition. As described in \prettyref{sec:Data-collection-and},
we gathered data from various sources in hope of making some confident
predictions.

The rest of the paper is structured as follows. We review some of
the previous work on the topic, ranging from specific analysis of
the ESC voting network to the more general methods of examining the
data. Then we present the dataset we worked with and how it was obtained.
We also present some of its main characteristics. Then we describe
in more detail the methodology used. Since some of the needed data
turned out to be very difficult to get hold of, some compromises had
to be made and these are discussed as well. In the last part we present
the results of the project and we end by brainstorming some future
work ideas and concluding the paper in a summary.

\section{Related work\label{sec:Related-work}}

Since we mainly deal with analysis of the ESC voting network, we discuss
some previous papers covering the topic in terms of applying network
techniques on the problem, introducing some ideas and techniques that
will prove useful for our project. Although they all deal with the
competition as a network problem to some extent, none of them use
more advanced network analysis tools such as community detection and
link prediction on the graphs, which we implement. Both these methods
can then be used for future projections, which is also not dealt with
in any of the papers.

\emph{\textcite{Mantzaris2018}} investigate different possible explanations
for the voting patterns which deviate significantly from a uniform
distribution, specifically focusing on the notion that nations try
to build reciprocal voting connections that lead to them receiving
more points from their ``partners'', and thus ranking higher. Therefore,
they try to find correlation between the number of collusive edges
a nation has and their success in the competition. They build on previous
work in (\textcite{alex2017examining,gatherer}), analyzing the voting
behavior by simulation voting, since analytical identification of
statistically significant trends in the competition would be mathematically
too complex because of its changing nature. Capturing the different
voting systems in place throughout the years mathematically is untraceable,
therefore simulation provides a good compromise.

The authors extend the algorithm presented in (\textcite{gatherer})
for finding significant exchange of points awarded between participants.
The original paper focused on a limited interval of competitions when
the voting rules were mostly homogeneous, therefore \emph{Mantzaris
et al. }provide a more general sampling technique. To be able to do
that, they identify the three principles of voting used by ECS since
its start in 1956. These can be grouped as \emph{allocated},\emph{
sequential }and \emph{rated.} The algorithm samples the uniform distribution
of points throughout a time period, based on the rules in place at
the time and then extracts the highest-weighted edges. Network is
formed based on those colluding edges between countries, showing patterns
of biased voting. They then perform community detection on obtained
structures and base their results on those. They consider both one-way
and two-way relationships and thus lay groundwork for thorough network
inspection in terms of both motif and community detection.

They find significant patterns of both preference and neglect spanning
throughout the participating nations, showing that voting is geographically
influenced, linking it to mutual history, similar ethnic features
and the feeling of ``brotherhood'' of neighboring countries. They
also conclude that the participants with higher number of colluding
edges achieve better success in the competition, showing it does pay
off to build partnerships. This, together with the changing nature
of the network that is more and more concentrated around the colluding
edges, implies that nations are actively trying to build these relationships.

\textcite{dekker} provides a different take on the analysis of the
voting network. The techniques the authors demonstrate have a more
general applicability, spanning away from the ESC, and can also be
used for analyzing other types of friendship networks. They focus
on the votes from the 2005 rendition of the contest and come up with
ways to adjust votes for song quality. With that, they produce a friendship
network with valued links (the value of the link being the strength
of the friendship). They find that friendships are often not returned,
which reveals their asymmetric nature, especially visible in countries
with a large number of immigrants.

They run a more statistical analysis by removing the influence of
song quality or popularity and it shows that friendship between countries
is determined in a big part by geographical proximity. Another factor
they find are large immigrant groups voting for their home country.
Other factors, such as population size, language similarity and economy
were found to be insignificant. They expose a visible five-bloc structure,
the blocs being the Eastern (former USSR countries, together with
Romania, Hungary and Poland), Nordic (Norway, Sweden, Denmark, Finland
and Iceland), Balkan (former Yugoslavia and Albania), Eastern Mediterranean
(Greece, Cyprus, Malta, Bulgaria and Turkey) and Western (Portugal,
Spain, Ireland, Andorra, Israel, the UK, France, Monaco, Germany,
Belgium and the Netherlands). Preferences among the different blocs
are also analyzed, finding that some blocs are more connected than
others. Grouping countries computationally by exposing the strongly
connected components, they find three different blocs. Using taxonomic
trees proved to be ineffective and only finding one bloc.

\textcite{escpolitical} analyze 29 years of the Eurovison Song Contest,
specifically the competitions held between 1975 and 2003. Its main
goal is to find any correlations between the points awarded and country
similarity, performance type, etc. The authors find some meaningful
properties impacting the scores and extract some clusters that exchange
votes regularly. They propose what could lead to this behavior, stating
that there exist cliques of countries that award points among themselves
and even trading with votes. But these blocs are found not to base
on politics, but rather on language and cultural similarities. To
measure the language impact, they rely on the Morris-Swadesh method
for analyzing linguistic differences.

To infer the influence of each factor, \emph{Ginsburgh et al.} formulate
a weighted expression, for which weights are assigned based on the
voting behavior. The major takeaway of it is that the biggest factor
influencing the voting decision is still the music quality. As with
the previously discussed work, they also notice an important role
of immigrants that vote for their country of origin. These observations
are, however, not algorithmic but rather the results of looking at
the formed communities and discussing the prevailing similarities
in them.

\section{Methods\label{subsec:Methodology}}

\subsection{Bias detection}

In the first major goal of the paper is determining the community
structure of the voting networks. The most important step is the formation
of edges that reflect a consistent bias between nations (both in terms
of positive and negative relationships) and we approach that in two
ways, described in this section. Both methods are used to detect bias
over all the selected time periods. Altogether, this gives us more
than 4400 different networks which are later used to present some
statistical facts about the distribution of points.

Firstly, we follow the methodology described in (\textcite{alex2017examining,Mantzaris2018,gatherer})
and used the Gatherer algorithm. This turned out to be the most effective
method and very important for our analysis, thus, we describe the
pseudo code in Algorithm \prettyref{alg:The-Gatherer-algorithm} and
Algorithm \prettyref{alg:Gathered-graphs}. Its main idea is to estimate
the number of points that a participant is expected to receive in
a certain time period, based on the rules in place at that time. It
then uses these estimates to find bias, i.e. behavior where nations
exchange more than the expected number of points in a time period.

The other method of obtaining the structure was developed by us and
it accounts for the number of points a country has received each year
in the selected period. We create a directed edge from country 1 to
country 2 if the first one awarded the second one more than the average
number of points received by the second one in more than 75 \% of
the competitions in that period. If the bias is shown both ways, we
add the edge to the undirected network. The pseudo code is described
in Algorithm \prettyref{alg:Average}.

We have generally found that the simulated voting implemented by the
Gatherer algorithm gives clearer and less noisy results. It proves
much more useful for detecting neglect, since the average points method
generates too much noise. Even the graphs generated by the Gatherer
algorithm were tricky to work with, which is why we added an additional
criteria to detect a neglect between 2 countries. Since geographic
proximity proved to be very important, we also demand that two neglecting
countries lie no more than 3 hops (borders) away on the map thus reducing
the amount of random edges.

Since the Gatherer algorithm is also used in (\textcite{alex2017examining,Mantzaris2018,gatherer}),
it is well tested and reliable. Therefore, we focus mainly on its
results for graph formation from here on. Although we were able to
extract some valuable information with the second method and it performed
very similarly to the Gatherer algorithm for the longer periods, it
behaves inconsistently on the shorter time spans, picking up too much
randomness, while the Gatherer algorithm performs consistently no
matter the period length, which led us to this decision.

We form two types of graphs: undirected, showing mutual affinity between
contestants (i.e. an undirected edge between two nodes is added if
both show bias towards each other), and directed, showing only one-way
bias. Here, we are more interested in actual one-way relationships
- an edge was therefore added only if one country shows bias towards
the other but the other does not show any bias for the first. Both
graphs use weighted edges, the weight denoting the difference between
the actual and expected (simulated / average) number of votes.

Despite the consistent performance by the Gatherer algorithm, it still
needs some tuning. Besides the noise picked up in neglect detection,
it also struggles on the directed networks, adding insignificant edges.
This is why we also post-process the directed graphs and remove the
edges whose weights were below the average in the network, giving
us much more readable results.

\subsection{Basic graph features}

In order to get the general oversight over the voting networks, we
calculate some basic statistics about the voting and bias networks,
such as the average number of nodes in a certain time period, the
average number of edges and the average degree. All the methods are
already implemented in the \emph{networkx }library (\textcite{NetworkX_library}).

\subsection{Community detection}

After extracting the biased voting trends, we extract the communities
using the \emph{Louvain} (\textcite{louvian}) community detection
algorithm in the undirected and the \emph{Newman's leading eigenvector
method }(\textcite{eigenvectors}) in the directed ones. Both are
implemented in the CDLib library (\textcite{cdlib}). The extracted
communities in the undirected network depict blocs of countries ``collaborating''
in the competition. The directed networks are analyzed somewhat differently,
since the actual communities do not play such a vital role here, as
there is no mutual point exchange. However, they still expose some
interesting behavior that would be missed if we only focused on the
undirected networks. The results of both types are presented in \prettyref{sec:Results}.

The number of extracted graphs also allows us to find the most commonly
co-occurring nations in communities. Those were extracted with the
\emph{apriori} algorithm, implemented in MLxtend library (\textcite{MLxtend}).

The plots seen in the Appendix B were generated with the Gephi visualization
tool (\textcite{gephi}).

\subsection{Correlation between the community structure and success in the competition}

One of the main goals of bias edge construction and community detection
was determining the affect the bias behavior has on the final score
of participants. In other words, we wanted to find out whether being
in a large community or having many friends in the network pays off.
Thus, based on the communities a node (country) belongs to, we gather
some voting data (points, points from community, percentage of points
from community and final place) and aggregate it for each community
type and period length.

We find the most interesting aggregations to be points per degree
in community, portion of points received from communities and total
number of points received from communities. Therefore we decided to
interpret those more carefully in \prettyref{sec:Results}.

\subsection{Preference detection and future projections}

One of the hypothesis we set was that users and jury vote based on
three main factors: the song popularity and features, the country
of the performance and the artist features. We split those categories
to subcategories and obtain as much data as possible about them. We
build a knowledge graph which connects all possible properties that
can be considered together into relations of different types. With
this graph we perform similarity scoring and link prediction where
we try to predict the ``voting relations'' based on other connections.
We use different link prediction algorithms which have to consider
the rich structure of the formed graph. In addition to network analysis
techniques, the dataset was also examined with the Orange package
\parencite{JMLR:demsar13a}.

\subsection{Prediction performance evaluation\label{subsec:Prediction-performance-evaluatio}}

It the second part of the analysis, we focus on the predictor of the
success in the competition. The performance is measured against the
performance of the betting tables. We use two different scores to
calculate the success of the model. The first score is the mean absolute
error (MAE) of the ranks inferred by the predictor based on the actual
results and the second score the \emph{recall at n }(\emph{Recall@n})
score for \emph{n =} \emph{3}, \emph{5} and \emph{10}. MAE, too, is
measured at distinct intervals: for the whole set of performing nations
and just for the top 10 performances each year.

\section{Data collection and presentation\label{sec:Data-collection-and}}

\subsection{Collection}

The data set used was obtained by scraping various web pages. The
voting data was collected from (\citetitle{ebu}) and the information
about specific countries, songs and performers was downloaded from
(\citetitle{eschome,wiki}). The available voting data includes all
points awarded by every country to every other participant throughout
the years, both for the final and the semi-finals (when both were
held), with the exception of the first ever competition in 1956, since
the data is not available. For the period between the years 2016 and
2019 we even got separated votes from jury and audience, since this
is when the EBU started sharing these figures.

For most countries we obtained their names, Wikipedia category entries,
languages, the currency, calling code, ethnic groups, religions, neighborhood
and some other features that we hope to be useful. For the participants
we have their country of origin, how old they were when the represented
their nation, name, Wikipedia categories, music genres, instruments
and occupations. Data for songs was scrapped from Wikipedia. We have
among other things the genres, categories, languages and released
date. To analyze songs even better we scrapped lyrics, chords and
scores from (\citetitle{lyrics_fandom,Musixmatch,musescore,ultimate_guitar}).
The biggest challenge presented the data about songs, performers and
performances themselves. We have tried to obtain as much as we could
from Wikipedia, at least for the latest entries, which were better
represented. We therefore focus mainly on those. We also plan to extract
some other important properties (the tones, harmony, metrum, melody...)
about song quality from the chords and scores and the prevailing themes
and motifs with text mining. Those features will be useful to pinpoint
the preferences of specific countries and the factors that contribute
most to success.

We have also obtained the betting tables for each competition between
the years 2004 and 2019 (\citetitle{odds}). These allow us to combine
our models with the expected outcomes based on the betting odds.

Wikipedia was therefore the main source of the data about the performers,
countries and music. This data is however not complete and we had
to make some compromises here, as discussed in \prettyref{sec:Problems-and-compromises}.

The data is stored in structured JSOG format (extended JSON format
which can work with references and is therefore better for graphs).
The scraping was done in Java but data analysis is done in Python
because of its numerous robust libraries for data management.

\subsection{The inferred networks}

Based on the collected voting information, we are able to form a large
number of graphs, showing the voting behavior throughout the history
of the competition.

Firstly, we just create the voting network for each contest separately
and for all of them together (the all-time voting network). The networks
for each competition are directed and any edge between two countries
depict the number of points awarded by one to another in that year.
The in-strength of any node therefore shows its total score that year.
Similarly, the all-time directed network depicts the total number
of points awarded in the competition history.

These networks are then used to form the bias networks as described
in section \prettyref{subsec:Methodology}. To observe the changes
throughout the competition history, we opt to form networks that represent
biases and neglect in certain periods - those were chosen to be 1,
5, 10, 15, 20, 25, 30, 35, 40, 45, 50, 60, 63 years. For each period,
both directed and undirected networks are created. This gives us more
than 4400 networks altogether, but we do not need to analyze all thoroughly.
The main focus are the networks that show the all-time preferences
(period length 63 years), the ones that depict different 10 year periods,
since this can show any changing nature of the voting, and the ones
that depict the last 20, 25 and 30 year periods, showing long-term
but still recent trends.

The all-time voting network has 52 nodes, one for each country that
has ever competed. The 10 most successful (the nations with the highest
number of points collected throughout the history) were Sweden, Norway,
the UK, Germany, France, Spain, Denmark, Greece, the Netherlands and
Ireland. The least successful so far have been Monaco, Bulgaria, Australia,
San Marino, Montenegro, Czech Republic, Slovakia, Andorra and Morocco.
However, these scores should not be too surprising and taken too seriously,
since the most successful nations are also the ones that have participated
in the competition the longest and many of the least successful ones
have only taken part a few times. On the other hand, Australia has
only participated 5 times so far and has achieved great success each
time, which can not be captured with this kind of analysis.

\subsection{Betting tables accuracy}

The baseline for measuring the performance of our prediction model
was using the betting tables as the only means for predicting the
outcome of the competition. Thus, this baseline needed to be determined.
The results were obtained using the performance metrics described
in \prettyref{subsec:Prediction-performance-evaluatio}, averaged
over the whole period for which we have obtained the betting tables.
They are presented in \prettyref{tab:Performance-evaluation-of-betting-tables}.
We can see the MAE of the tables improves for the higher part of the
table, while the recall does not seem to be affected much by the range.
These figures are the baseline for our model, which provided results
described in \prettyref{subsec:Prediction-results}.

\begin{table}
\begin{tabular}{|c|c|}
\hline 
Performance measure used & Results\tabularnewline
\hline 
\hline 
Mean averaged error over the whole set & 4.3391\tabularnewline
\hline 
Mean averaged error over the top 10 & 4.0421\tabularnewline
\hline 
Recall@3 & 0.4386\tabularnewline
\hline 
Recall@5 & 0.54737\tabularnewline
\hline 
Recall@10 & 0.56842\tabularnewline
\hline 
\end{tabular}

\caption{Performance evaluation of betting tables as predictors\label{tab:Performance-evaluation-of-betting-tables}}
\end{table}

\section{Results\label{sec:Results}}

The results are grouped into multiple subsections, dealing with the
communities of positive bias, countries showing neglect, the correlation
between the community structure of a country and its success, what
we think causes this behavior, the inferred preferences of specific
countries and the prediction results.

\subsection{Communities}

The number of generated networks makes it possible to reason about
the different trends and influences on the voting. Although we could
have focused on any period in the competition history, we chose to
further inspect the most recent results and mostly summarize the older.

\prettyref{fig:Bidirectional-bias-from-1956} shows the communities
formed if we consider the results from the start of the competition
in 1956. There are 10 communities in total and they are strongly geographically
influenced, forming the following blocs: Northern (Sweden, Denmark,
Norway, Iceland), Western (Ireland, United Kingdom, Germany, Luxembourg),
Southern (Italy, Malta, Spain, Portugal), Central (Netherlands, Hungary,
Belgium, Austria), Baltic (Lithuania, Estonia, Latvia, Finland), Eastern
(Poland, Ukraine, Russia, Belarus), the Balkan (Greece, Cyprus, Albania),
South-Western (Moldova, Romania, Turkey), Yugoslavian (Croatia, Slovenia)
and Cross-Continental (Israel, France). Especially prominent are the
connections between Cyprus and Greece, Greece and Albania, Romania
and Moldova, Italy and Malta and the former USSR countries.

Although this is the network that includes the most data and is thus
seemingly the most important, we only mention it here for the sake
of completeness. We are more interested in the networks depicted in
\prettyref{fig:Bidirectional-bias-from-1999}, \prettyref{fig:Bidirectional-bias-throughout}
and \prettyref{fig:Bidirectional-bias-10} for the later parts of
the analysis as they speak of the more recent trends.

\prettyref{fig:Bidirectional-bias-throughout} and \prettyref{fig:Bidirectional-bias-10}
show how the bias networks have evolved and grown, although the main
communities remain the same. Clearly, there is more and more biased
voting, but it remains concentrated in the same blocs in all periods.
It is interesting to see how Australia got mixed into the Northern
bloc in the last 10 years. This may be one of the reasons for their
reasonable success so far. During the first five time they have taken
part in the competition, they showed a very focused voting behavior
and at the same time managed to collect many points from until then
a very closed bloc.

We think the most interesting and current network is the one in \prettyref{fig:Bidirectional-bias-from-1999},
since it shows the recent trends, while still taking into account
a longer time period. The communities are very similar to the ones
implied by the all-time bias network, showing the persistence of these
relationships.

Also interesting is the network shown in \prettyref{fig:Unidirectional-bias-from-1956},
showing the all-time network of one-way relationships. The edges depict
relationships where only one country awards more than average number
of points and the other does not. Communities are not that prominent
in this network, but still visible. One reason why the community structure
is limited in the fact that historically very successful countries
such as Sweden receive a high number of points from others very often
and they can not ``return'' the votes to all of them. Therefore,
they have a very high in-degree and this does not infer any preference,
just the fact that they were successful. However, some relationships
in the all-time network are still quite interesting, like the strong
edge from Croatia to Bosnia and Herzegovina and the edges from the
former USSR nations to Russia.

The data also allows us to find sets of countries that end up in the
same communities most often. The results are presented in \prettyref{tab:Countries-that-ended}.
As the table shows, the countries that co-occur in a community most
often are Cyprus and Greece, which are a part of more than 10 \% of
the formed communities. They are followed by some Scandinavian countries
and the most regular participants in the competitions, such as the
UK, Ireland and Switzerland. The most common set of size three contains
Denmark, Sweden and Norway. We also notice a strong relationship between
Portugal and Spain, Romania and Moldova, Slovenia and Croatia and,
interestingly, France and Israel. All the relationships are also visible
in the figures in \prettyref{sec:Voting-networks}.

\begin{table}
\begin{tabular}{|c|c|c|}
\hline 
Rank  & Countries  & Relative support\tabularnewline
\hline 
\hline 
1  & Cyprus, Greece  & 0.108919\tabularnewline
\hline 
2  & Denmark, Sweden  & 0.080505\tabularnewline
\hline 
3  & Sweden, Norway  & 0.069455\tabularnewline
\hline 
4  & Switzerland, United Kingdom  & 0.065904\tabularnewline
\hline 
5  & Denmark, Norway  & 0.062352\tabularnewline
\hline 
6  & United Kingdom, Ireland  & 0.061957\tabularnewline
\hline 
7  & Denmark, Sweden, Norway  & 0.055249\tabularnewline
\hline 
8  & Spain, Portugal  & 0.053275\tabularnewline
\hline 
9  & Sweden, Iceland  & 0.042620\tabularnewline
\hline 
10  & Germany, United Kingdom  & 0.041831\tabularnewline
\hline 
11  & Denmark, Iceland  & 0.041436\tabularnewline
\hline 
12  & Romania, Moldova  & 0.036701\tabularnewline
\hline 
13  & Belgium, Netherlands  & 0.036306\tabularnewline
\hline 
14  & Slovenia, Croatia  & 0.034728\tabularnewline
\hline 
15  & Israel, France  & 0.034333\tabularnewline
\hline 
16  & Denmark, Sweden, Iceland  & 0.033149\tabularnewline
\hline 
17  & Norway, Iceland  & 0.033149\tabularnewline
\hline 
18  & Germany, Ireland  & 0.032755\tabularnewline
\hline 
19  & Finland, Sweden  & 0.028808\tabularnewline
\hline 
20  & Estonia, Latvia  & 0.027624\tabularnewline
\hline 
\end{tabular}

\caption{Countries that ended up in the same community most often and the relative
number of times\label{tab:Countries-that-ended}}
\end{table}

\begin{figure*}
\includegraphics[width=1\textwidth]{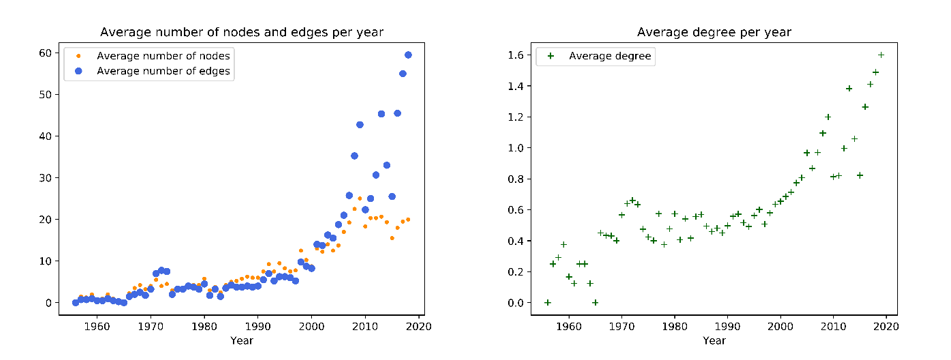}

\caption{Average number of nodes and edges, average degree and clustering coefficient
in the bias networks throughout the years.\label{fig:Average-number-of}}
\end{figure*}

The graphs in \prettyref{fig:Bidirectional-bias-throughout} and \prettyref{fig:Bidirectional-bias-10}
provide a different view as to how the bias has evolved throughout
the history and it is clear that there are more and more biased connections.
This can also be seen if we look at the average degree of the bias
undirected network throughout the history, depicted in \prettyref{fig:Average-number-of}.
The degree has been rising consistently, which means that the countries
are actively forming more and more friendship communities and concentrating
their votes among specific ``partners''.

\subsection{Correlation between the community structure and success}

\prettyref{tab:AVGpercentagePointsFromCluster} shows what percentage
of points countries got from their communities. For example, averaged
over 25 years, members of neglect communities only received 6.1 \%
of their points from that community, while members of communities
in directed graphs get on average 17.6 \% of their points from that
cluster.

\prettyref{tab:AVGPlace}, \prettyref{tab:AVGpoints} and \prettyref{fig:Degree-points-50}
show a recurring relationship between the success of a nation and
its community structure. Being in a positive bias community pays off
because countries get on average higher scores and achieve higher
places, a trend clearly visible in the plots. We can also see that
it is better to avoid neglect clusters, since membership in those
usually means lower ranking.

\subsection{Neglect}

We form the neglect networks in a similar manner to the positive bias
ones and the results are shown in \prettyref{fig:Bidirectional-neglect-from-1956},
\prettyref{fig:Bidirectional-neglect-30} and \prettyref{fig:Bidirectional-neglect-10}.
As expected, some distinct neglect relationships are visible between
nations, most notable between Macedonia and both Greece and Cyprus.
Similar holds for the pair Cyprus and Turkey and more recently, for
Azerbaijan and Armenia. Interestingly, there is also a strong evidence
of neglect between Germany and the pair Belarus and Ukraine. As seen
in the more recent networks, the trends persist.

\subsection{Possible influences and motivations}

As found in the discussed literature, geographical proximity seems
to influence the voting behavior most, as can be seen through the
geographically local communities that form. Moreover, affinity between
nations such as the UK and Malta stress that language similarity also
plays a role. Common historical background could be attributed to
the affinity between the former Yugoslavian and USSR nations, since
the communities rarely extend beyond the bounds of the former unions.

The one way relationships are trickier and less obvious. However,
they can be explained to some extend by the number of immigrants (e.g.
votes from Croatia to Bosnia and Herzegovina, Germany and France to
Turkey and Switzerland to Serbia) and historical significance of one
country to the other (e.g. the votes from the former USSR countries
to Russia). Other reasoning is hard to ground since the highest in-degrees
can be explained purely on the success in the competition.

It is worth noticing that the positive bias behavior is most strongly
represented by pairs of nations that are more isolated, either geographically
(e.g. Spain and Portugal, the UK and Ireland, Romania and Moldova,
the Scandinavian countries), or culturally (e.g. Cyprus and Greece,
Greece and Albania, the Baltic countries and, again Romania and Moldova).

The countries showing most neglect have notable reasons as well. Especially
the historical relationship between Macedonia and Greece and more
recently between Albania and Serbia can be explained by their non-friendly
neighborhood relations.

\subsection{Nation's music preferences}

The constructed knowledge graph and Orange visualization tools offer
a glimpse into what genres, music styles and other performance features
caught the voters attention. Unfortunately, due to the lack of data,
we are only able to extract the crudest of relationships, thus, we
do not discuss them here thoroughly. Some trends we observe, though,
are the fondness of Slovenia towards Croatian songs (both in the form
of the language and the origin), Australia towards songs in English
and we again confirm the strong relationship between Greece and Cyprus.

\subsection{Prediction performance evaluation\label{subsec:Prediction-results}}

The performance measures indicate that the built model did not increase
the accuracy of the betting tables. Much of this can be attributed
to the fact that the data was often very sparse and not structured
very well. Even after preprocessing and filtering the whole dataset,
we were still left with too many unreliable and altogether not very
useful entries.

In \prettyref{tab:Performance-evaluation-of} we report the performance
measures when we also consider the predictor data together with the
data from the betting tables in variable amounts. The hyperparameter
$\beta$ indicates how much the predictions made by the our model
are taken into account ($\beta=0$ means only the betting tables are
used and $\beta=1$ means we rely only on our predictor). We can see,
the predictor does not improve the betting tables performance.

\begin{table*}[t]
\begin{tabular}{|c|c|c|c|}
\hline 
Performance measure used & Results ($\beta=0.2$) & Results ($\beta=0.5$) & Results ($\beta=1.0$)\tabularnewline
\hline 
\hline 
Mean averaged error over the whole set & 5.99516 & 6.52149 & 6.75597\tabularnewline
\hline 
Mean averaged error over the top 10 & 6.48421 & 7.46316 & 7.8\tabularnewline
\hline 
Recall@3 & 0.14035 & 0.08772 & 0.07018\tabularnewline
\hline 
Recall@5 & 0.23158 & 0.2 & 0.11579\tabularnewline
\hline 
Recall@10 & 0.42105 & 0.35263 & 0.31053\tabularnewline
\hline 
\end{tabular}

\caption{Performance evaluation of our prediction model\label{tab:Performance-evaluation-of}}
\end{table*}

\section{Problems and compromises\label{sec:Problems-and-compromises}}

The incompleteness of the data has turned out to be a problem very
early on, as we were initially unable to construct graphs based on
some similarities, namely the ethnic groups, immigrant numbers and
economic exchange. We thus resorted to manual inspection of the probable
causes of some trends. The inferred relationships are thus based only
on our domain knowledge and presumptions.

As expected, the availability of the data about the performances,
songs and authors is also limited, but we have managed to obtain a
reasonable amount of it and we hope it will prove useful for the second
part of the project.

Another problem we encountered was the noise in the less robust networks
such as the directed ones and the ones dealing with neglect. They
needed a lot of tuning and some post-processing to present any usable
information, but the final outcome is still quite non-deterministic
and open to numerous interpretations.

Motif counting and detection was also found to be not as effective
as we had hoped. The process of extracting the motif structure itself
was not very straight-forward since the functionality is not as widely
implemented as some other tools and at the same time the results were
not as informative and interpretable as the community structure itself.
For example, the notion of the reciprocal point exchange is summed
up in the undirected positive bias networks. Thus, we think that a
thorough inspection of the motif structure would not provide better
enough understanding of the network. We therefore abandoned this idea
and focused on other analysis tools.

If we were able to manage the dataset deficits in the first part of
the paper, they really came forward in the second part, since the
shortcomings disabled us to build a valid and useful model for prediction.
We leave this feat for future work.

\section{Future work}

During the analysis, we came across a few possible applications to
other fields. Firstly, the ideas and method discussed here do not
necessarily apply only on the Eurovision voting network. Such analysis
can be applied to any voting system, especially ones with a smaller
number of voting entities, such as the participating countries discussed
in this paper. We would find analysis of the voting behavior in sports
where points are awarded by judges from different countries very interesting.
These sports include ski jumping, figure skating, gymnastic etc. Similarly,
taking a closer look at the voting for awards would presumably reveal
interesting trends. One of such awards is the Ballon d'Or prize in
soccer, where journalists and players from around the world vote for
the best footballer each year. Each nation is represented by its journalists
and players, which is similar to the voting structure of the ESC.

A different field we would also be interested in is the voting a political
environment such as the European Parliament. As representatives from
the whole EU vote for propositions which come from different backgrounds,
one might find some trends in the way the representatives from specific
countries vote.

Lastly, we consider our own implementations and dataset. Some methods
we implemented did not take into account all the specifics in the
ESC dataset (e.g. the change of Macedonia to North Macedonia was handled
manually) and could be extended to further increase the result reliability.
One of the main objectives for future work would also be the aforementioned
expansion of the dataset that could allow a better model of the behavior. 

\section{Summary and conclusions}

In this paper, we analyzed the trends in the ESC voting network. The
results show strong and recurring patterns of mutual point exchange
between neighboring countries. We observed the most commonly recurring
friendships and one-way relationships together with some persistent
behavior of neglect. As discussed in the previous work, they can be
explained by geographical proximity and language similarity, as well
as ethnic structure and historical bonds. Having a large number of
biased relationships positively correlates to the success in the competition
and we observed more and more relationships the the more recent years.
Isolation of sets of countries seems to make bonds among the members
of the set stronger. 

We also described the methodology used in more detail and explained
how the data was structured to obtain the information. The obtained
data was then used to build predictor for future contests. To the
extent possible, we leveraged the distinct music preferences of individual
nations to extract which genres and music styles achieve the greatest
success in different countries. This involves both the points given
by the nation to other countries for their performances and also their
representative artists. This data was combined with betting tables,
since they are widely considered to be the best predictors about the
success of participants. The resulting model did not outperform the
betting tables alone with its main weakness being the lack of reliable
data.

We look forward to future extensions of our work on similar fields
or the same project with a more promising prediction model.

\printbibliography

\onecolumn

\appendix

\section{Pseudo-code\label{sec:Pseudo-code}}

\begin{algorithm}[H]
\begin{lyxcode}
\textbf{function}~Gatherer~(start\_year,~end\_year)~~~

conf\_up~=~bias~threshold~~//~1~\%~in~our~case

conf\_low~=~neglect~threshold~~//~~90~\%~in~our~case

avg\_simulation~=~{[}{]}

//~simulate~voting~enough~times~

//~to~obtain~a~reliable~expectation

//~(100000~times~in~our~case)~~~

\textbf{for}~selected~number~of~iterations:~~~~

simulation~=~{[}{]}

\textbf{for}~year~\textbf{in}~start\_year..end\_year:

~~~~score~=~expected~(uniformly~random)~

~~~~~~~~number~of~votes~received~by~a~contestant

~~~~//~depends~on~the~voting~scheme~~~

~~~~append(simulation,~score)

avg\_sim~=~mean(simulation)~~~~~~~~~

append(avg\_simulation,~avg\_sim)

sort(avg\_simulation,~reverse=True)

positive\_bias~=~percentile(avg\_simulation,~conf\_up)~~

//~more~than~bias~number~of~points~reflect~biased~voting

negative\_bias~=~percentile(avg\_simulation,~conf\_low)

//~more~than~bias~number~of~points~reflect~neglect
\end{lyxcode}
\caption{The Gatherer algorithm\label{alg:The-Gatherer-algorithm}}
\end{algorithm}

\begin{lyxcode}
\begin{algorithm}[H]
\begin{lyxcode}
function~Determine\_Bias\_Gatherer~(start\_year,~end\_year)~~

period\_length~=~start\_year~-~end\_year~+~1

participants~=~nations~that~took~part~in~the~ESC~in~the~period

for~c1~in~participants:~~~~~~~~~~~

for~c2~in~participants:~~

~if~times\_participating\_together~>~period\_length~/~5:~~~~

~//~only~take~into~account~the~participants~

~//~that~took~part~in~20~\%~of~all~competitions~in~that~period

~~~~~points\_awarded\_1,~points\_awarded\_2~=~

~~~~~~~~~number~of~points~awarded~by~c1~to~c2~(and~by~c2~to~c1)~in~the~period

~~~~~threshold\_high~=~the~threshold~number~of~points~for~that~period

~~~~~~~~~showing~bias~calculated~by~the~Gatherer~algorithm

~~~~~threshold\_low~=~the~threshold~number~of~points~for~that~period

~~~~~~~~~showing~neglect~calculated~by~the~Gatherer~algorithm

~~if~points\_awarded\_1~>~threshold\_high~>~points\_awarded\_2:~

~~~~~//~one-way~bias~

~~~~~add~edge~(c1,~c2)~with~weight~

~~~~~~~~~(points\_awarded\_1~-~points\_awarded\_2)~

~~~~~to~the~directed~bias~network

~~if~points\_awarded\_1~>~threshold\_high~and~

~~~~~points\_awarded\_2~>~threshold\_high:~

~~~~~//~two-way~bias

~~~~~add~edge~\{c1,~c2\}~with~weight~

~~~~~~~~((points\_awarded\_1~+~points\_awarded\_2)~/~2~-~~threshold\_high)~

~~~~~to~the~undirected~bias~network

~~if~c1~less~than~3~hops~away~from~c2:~~

~~~~~//~only~consider~countries~that~are~close~

~~~~~//~geographically~to~avoid~noise

~~~~~if~points\_awarded\_1~<~threshold\_low~

~~~~~~~~and~points\_awarded\_2~<~threshold\_low:~//~two-way~neglect

~~~~~~~~add~edge~\{c1,~c2\}~with~weight~

~~~~~~~~~~~(threshold\_low~-~(points\_awarded\_1~+~points\_awarded\_2)~/~2)~

~~~~~~~~to~the~undirected~neglect~network
\end{lyxcode}
\caption{Method of forming a bias voting network based on statistics calculated
by the Gatherer algorithm\label{alg:Gathered-graphs}}
\end{algorithm}

\begin{algorithm}[H]
\begin{lyxcode}
function~Determine\_Bias\_Average~(start\_year,~end\_year)~~

overshot~=~defaultdict(int)

period\_length~=~start\_year~-~end\_year~+~1

threshold~=~0.75~~

//~threshold~of~how~many~times~more~than~the~average~

//~number~of~points~need~to~be~awarded~for~bias~to~occur

participants~=~nations~that~took~part~in~the~ESC~in~the~period

for~year~in~start\_year..end\_year:~~

determine~the~average~number~of~points~

~~~~for~each~participant~in~the~time~period

for~c1~in~participants:~~

for~c2~in~participants:

~~~~points\_awarded\_1,~points\_awarded\_2~=~

~~~~~~~~number~of~points~awarded~by~c1~to~c2~(and~by~c2~to~c1)~in~the~period~

~~~~determine~how~many~times~each~country~has~awarded~any~other~more~than~

~~~~~~~~the~average~number~of~points~received~by~the~second~in~the~time~period

~~~~appearances\_1,~appearances\_2~=~

~~~~~~~~number~of~times~c1~(and~c2)~participated~in~the~ESC~in~the~period

~~~~overshot\_1,~overshot\_2~=~

~~~~~~~~number~of~times~c1~gave~more~than~the~average~number~of~points~

~~~~~~~~~~~~received~by~c2~to~c2~in~the~period

~~if~overshot\_1~>~threshold~{*}~appearances\_2~and~~

~~~~~overshot\_2~<~threshold~{*}~appearances\_1:~

~~~~~//~one-way~bias~

~~~~~add~edge~(c1,~c2)~with~weight~

~~~~~~~~~(overshot\_1~-~overshot\_2)~

~~~~~to~the~directed~bias~network

~~if~points\_awarded\_1~>~threshold~{*}~appearances\_2~and~

~~~~~points\_awarded\_2~>~threshold~{*}~appearances\_1:~

~~~~~//~two-way~bias

~~~~~add~edge~\{c1,~c2\}~with~weight~

~~~~~~~~~((overshot\_1~+~overshot\_2~-~

~~~~~~~~~~~~~threshold~{*}~appearances\_1~-~threshold~{*}~appearances\_2)~/~2)~

~~~~~to~the~undirected~bias~network
\end{lyxcode}
\caption{Method of forming a bias voting network based on the average number
of points received by the countries in the time period\label{alg:Average}}
\end{algorithm}
\end{lyxcode}

\section{Voting networks\label{sec:Voting-networks}}

\begin{figure}[H]
\includegraphics[width=1\linewidth,height=1\textheight,keepaspectratio]{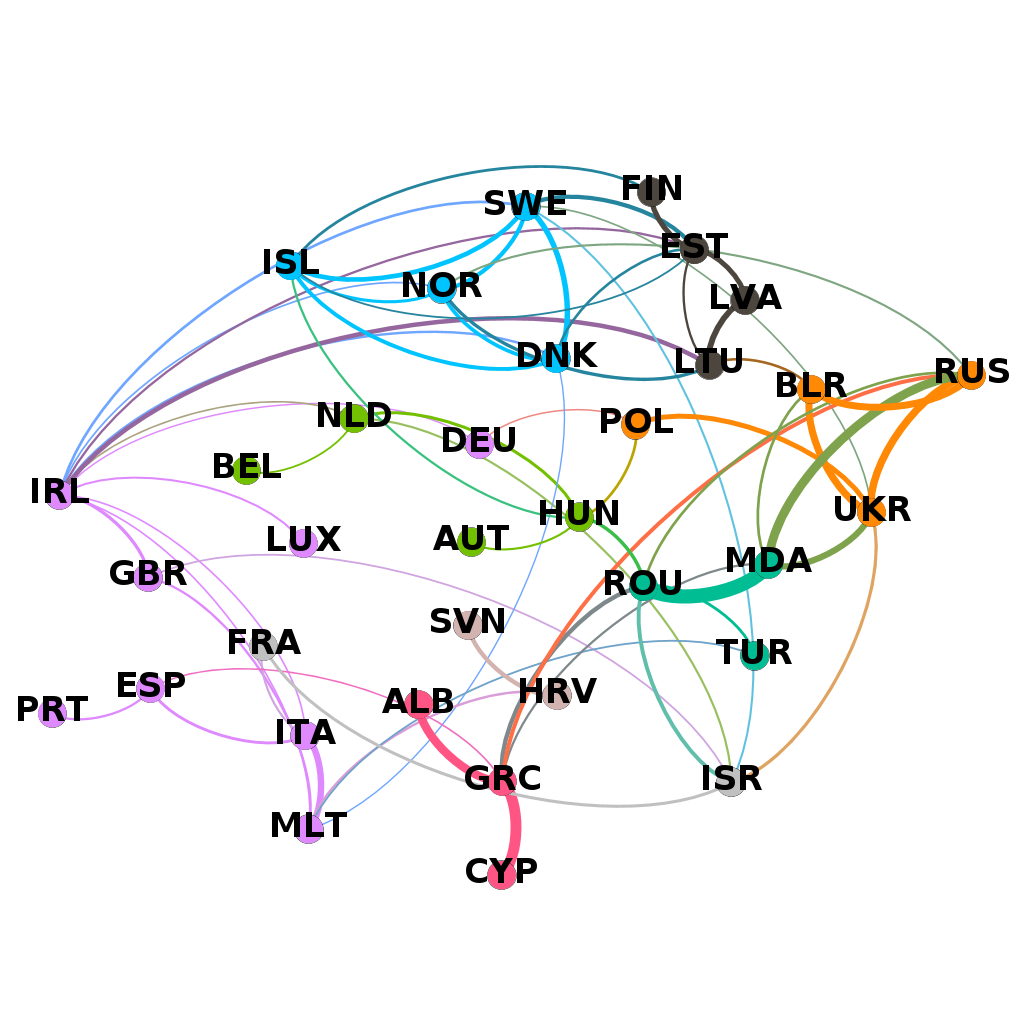}

\caption{Bidirectional bias from the start of the competition.\label{fig:Bidirectional-bias-from-1956}}
\end{figure}

\begin{figure}[H]
\includegraphics[width=1\textwidth,height=1\textheight,keepaspectratio]{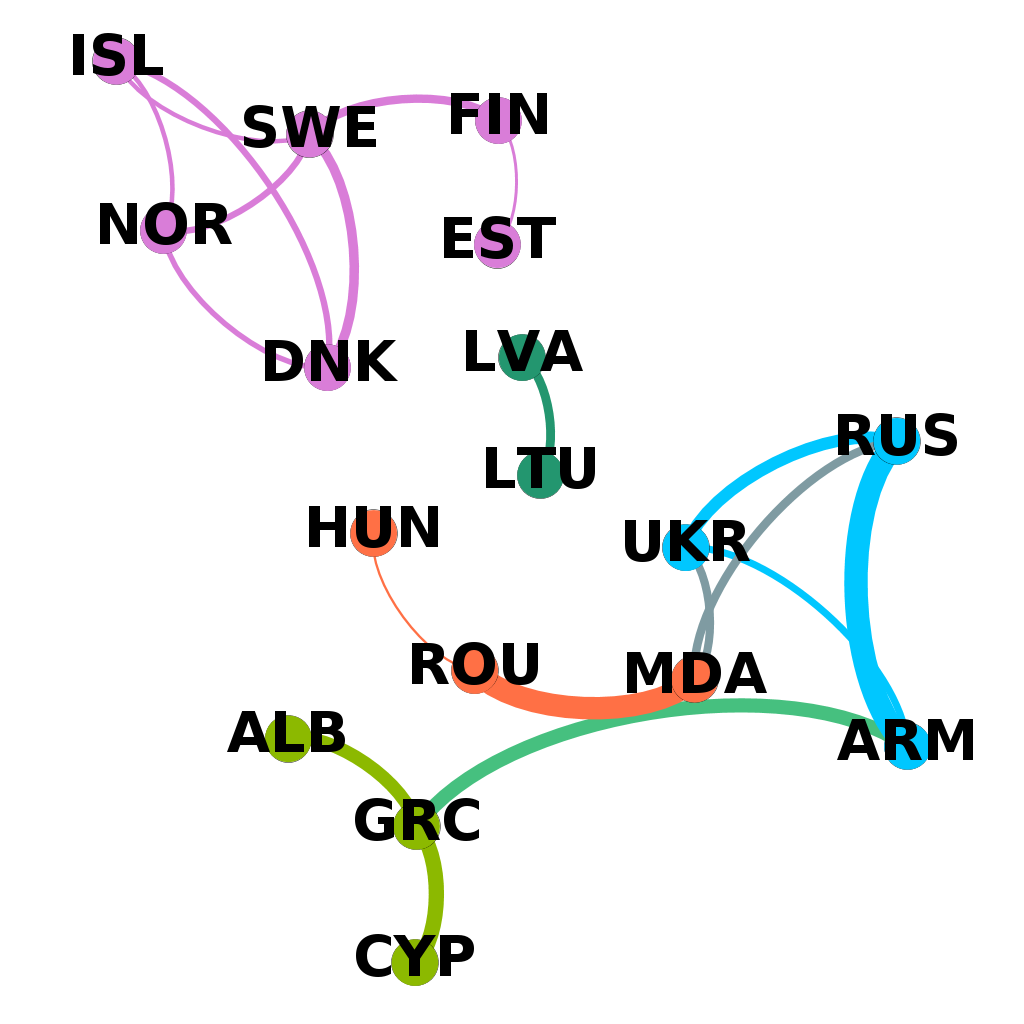}

\caption{Bidirectional bias from the last 20 years (1999-2019).\label{fig:Bidirectional-bias-from-1999}}
\end{figure}

\begin{figure}[H]
\includegraphics[width=0.2\textwidth,height=0.2\textheight,keepaspectratio]{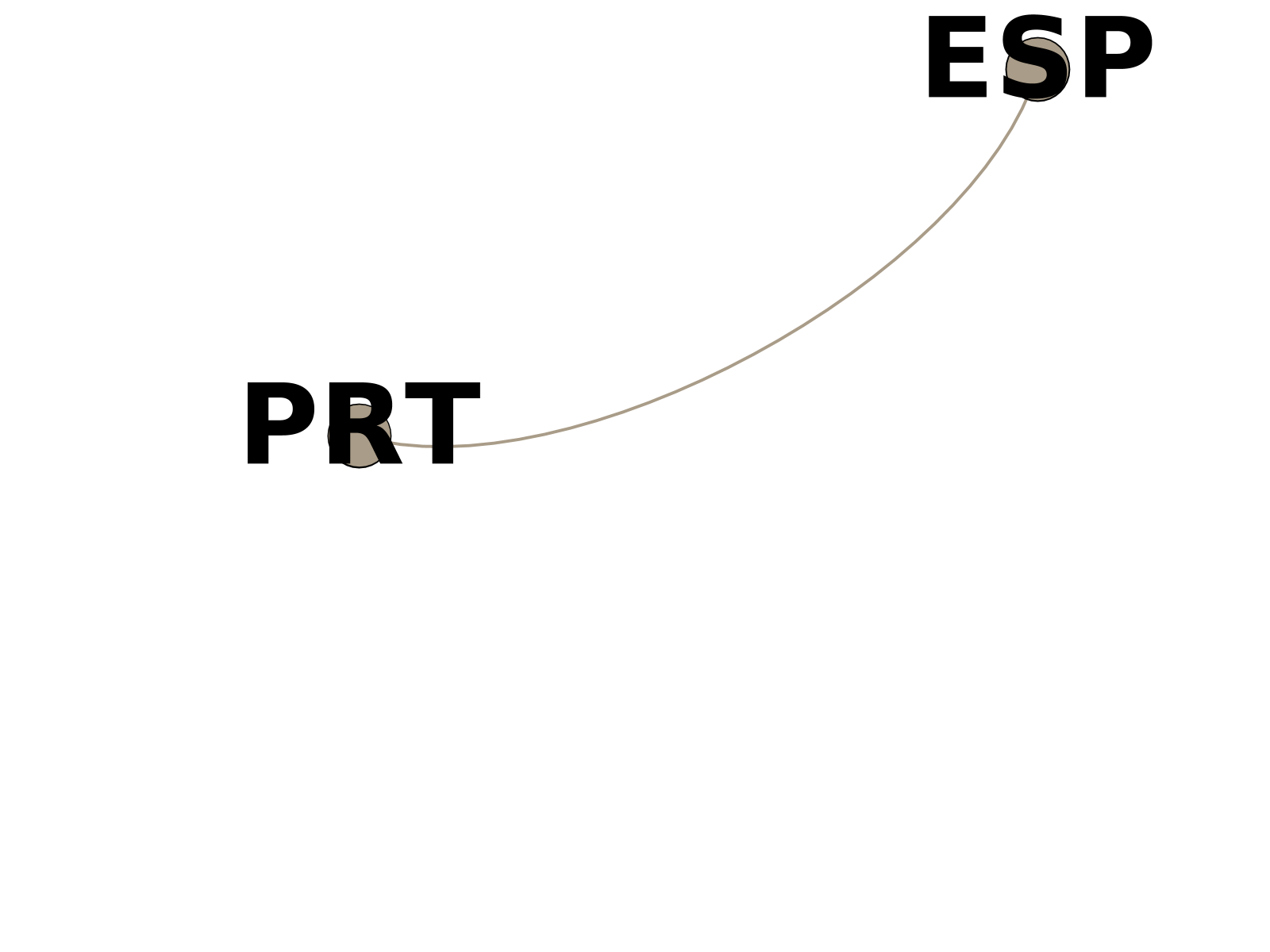}\includegraphics[width=0.2\paperwidth,height=0.2\textheight,keepaspectratio]{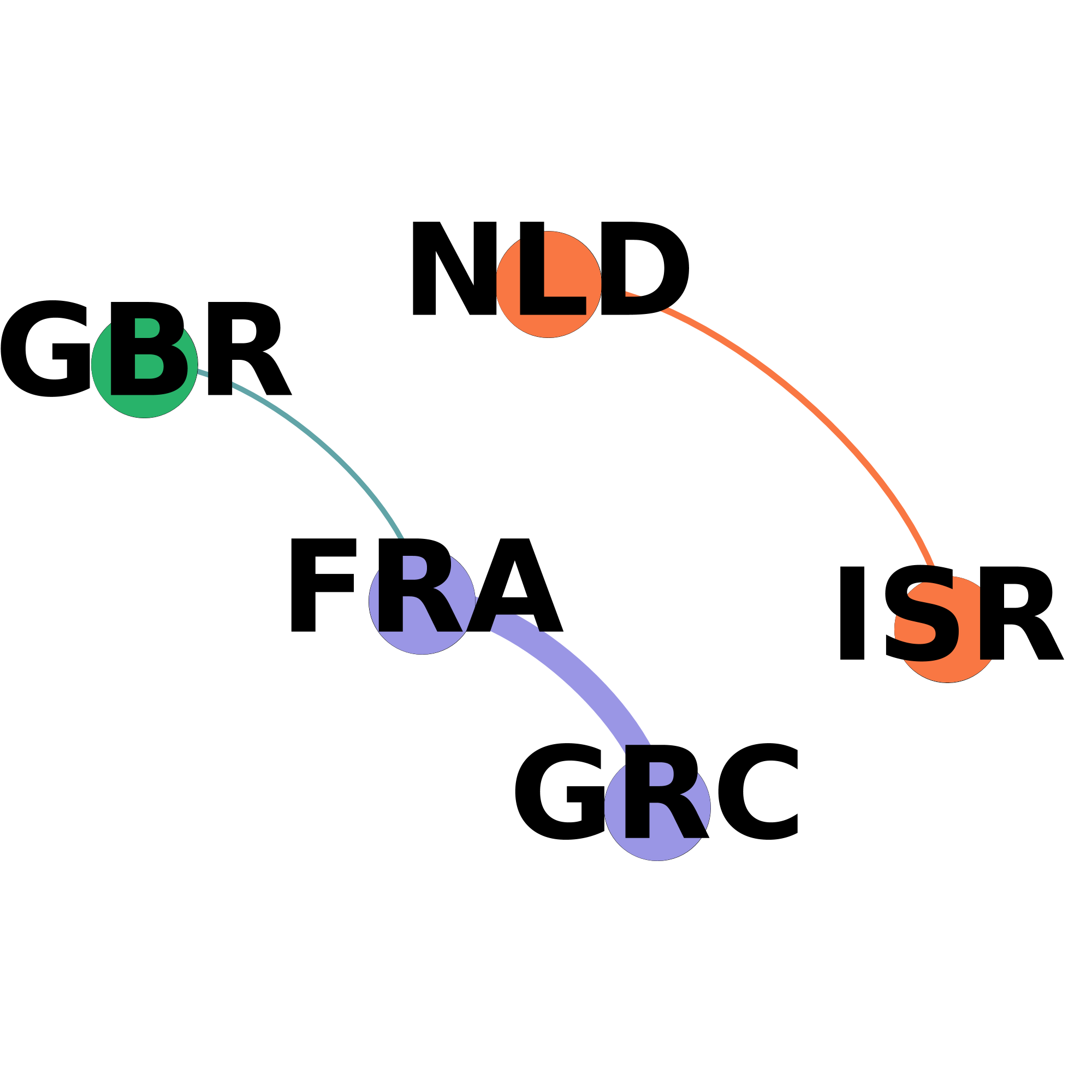}\includegraphics[width=0.2\paperwidth,height=0.2\textheight,keepaspectratio]{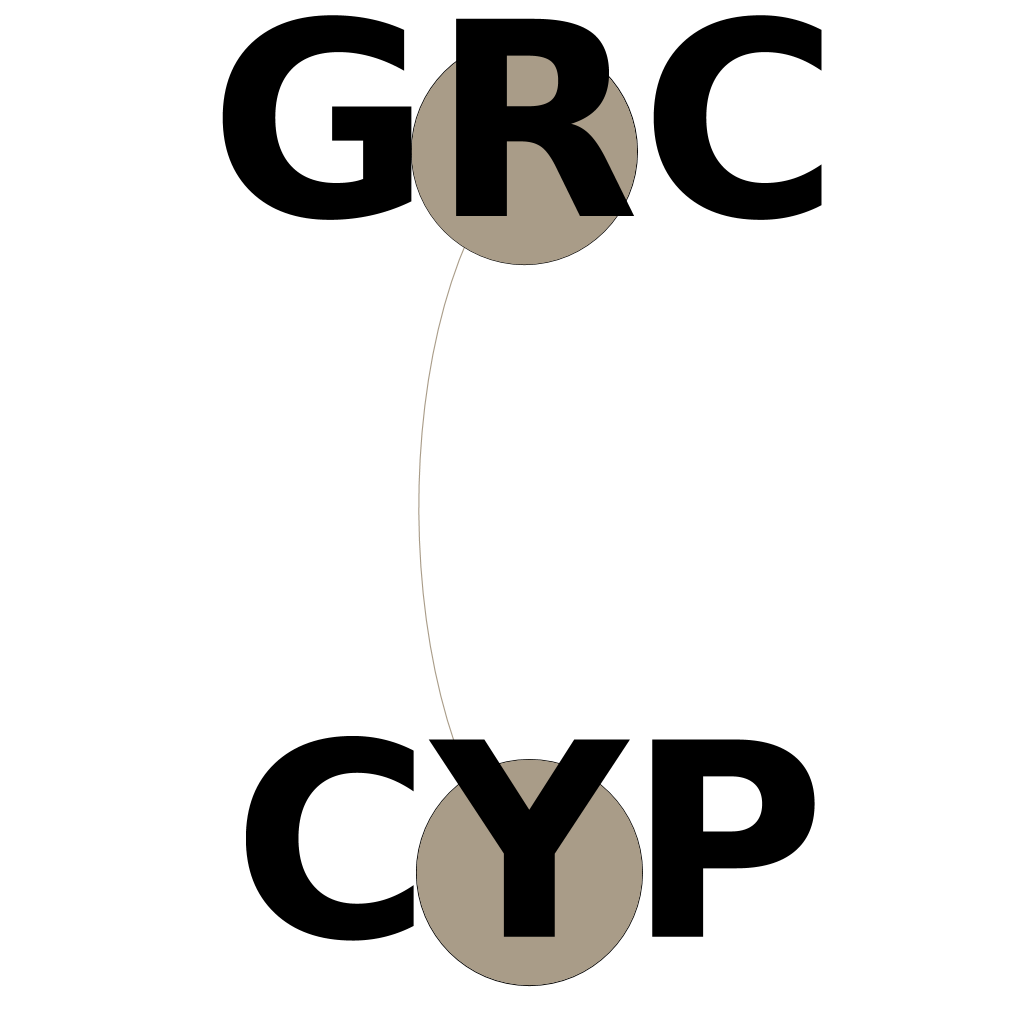}\includegraphics[width=0.2\paperwidth,height=0.2\textheight]{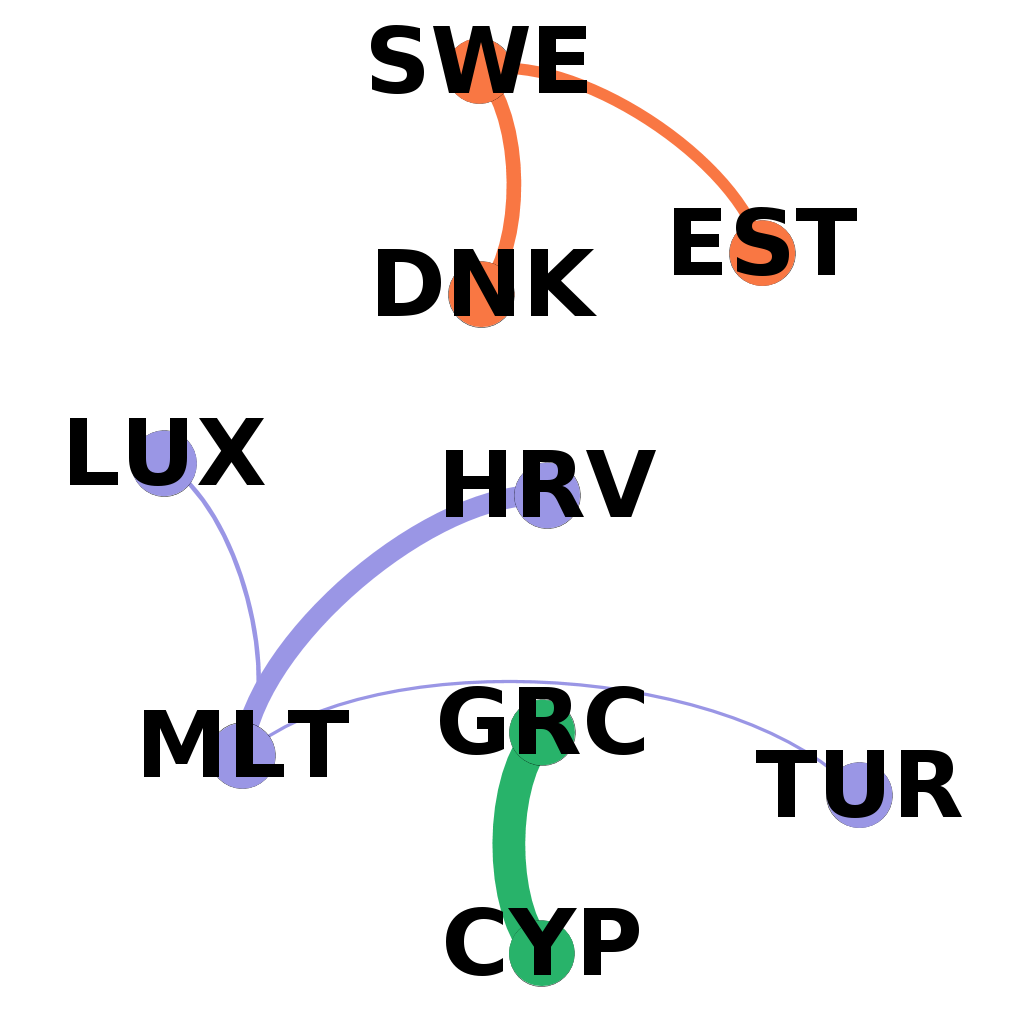}

\caption{Bidirectional bias in 10 year periods (1959-1969, 1969-1979, 1979-1989,
1989-1999).\label{fig:Bidirectional-bias-throughout}}
\end{figure}

\begin{figure}[H]
\includegraphics[width=0.5\textwidth,height=0.5\textheight,keepaspectratio]{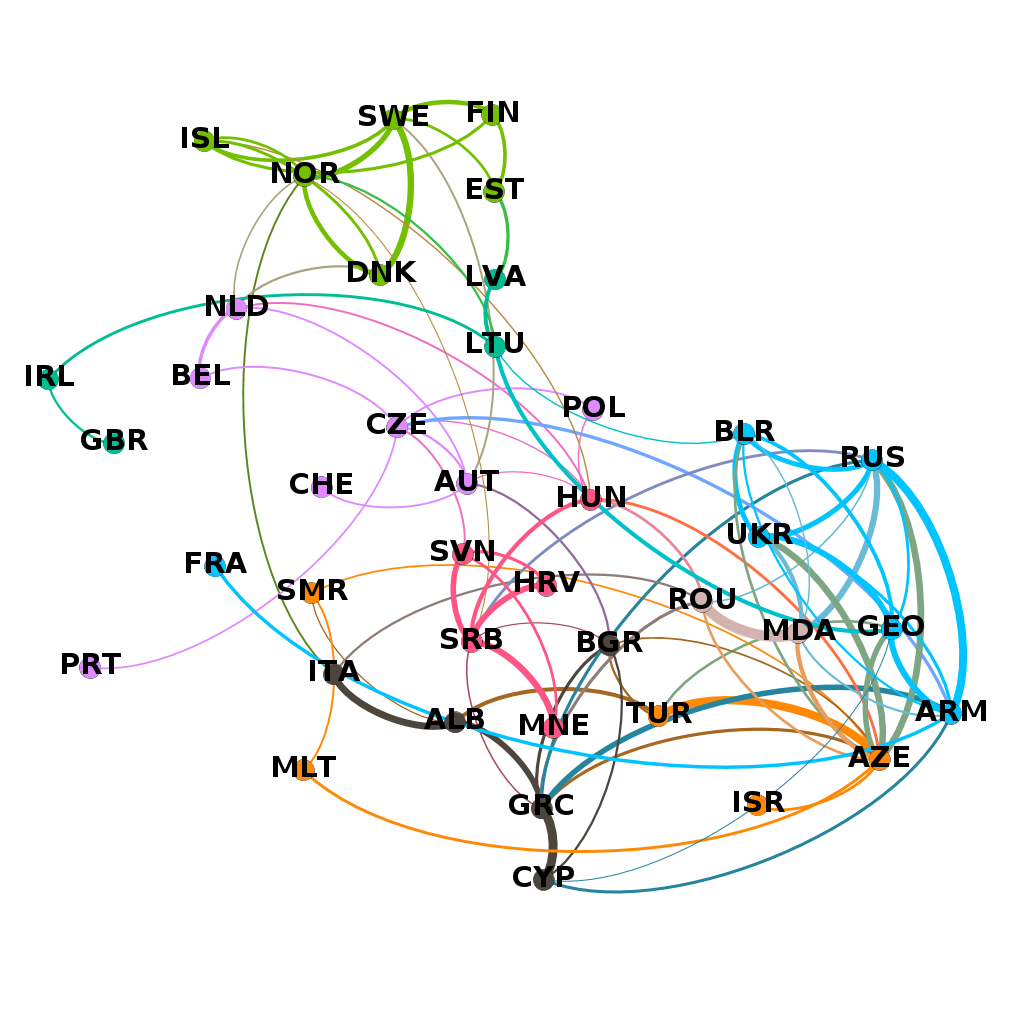}\includegraphics[width=0.5\textwidth,height=0.5\textheight,keepaspectratio]{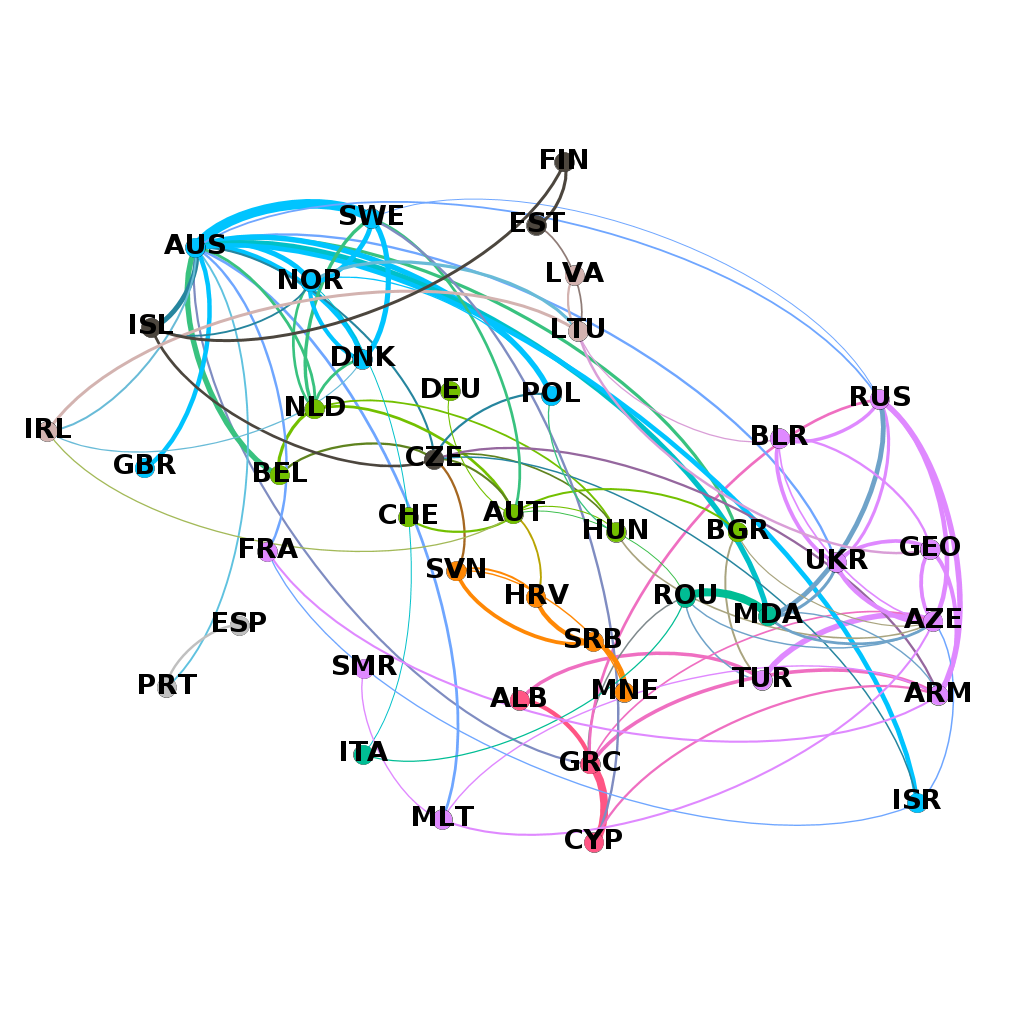}

\caption{Bidirectional bias in 10 year periods (1999-2009, 2009-2019).\label{fig:Bidirectional-bias-10}}
\end{figure}

\begin{figure}[H]
\includegraphics[width=14cm,height=14cm,keepaspectratio]{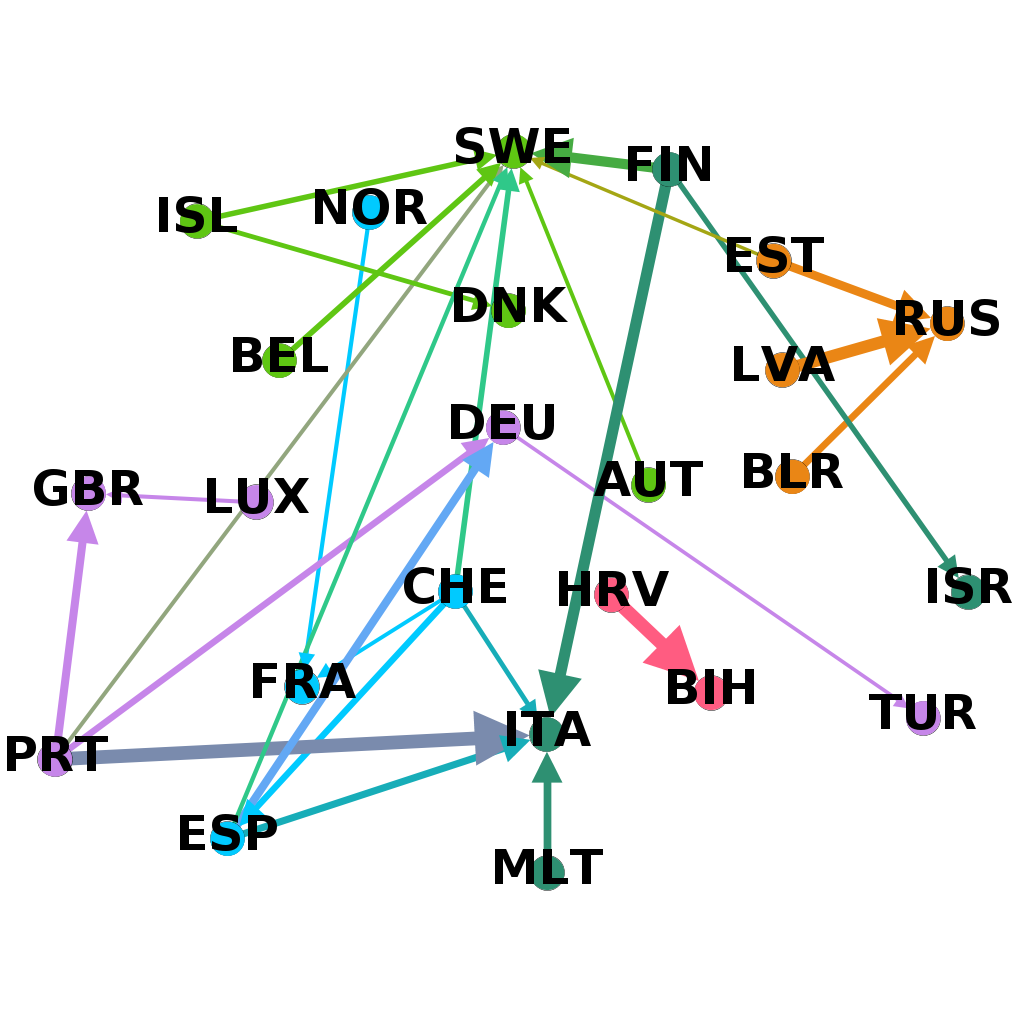}

\caption{Unidirectional bias from the start of the competition.\label{fig:Unidirectional-bias-from-1956}}
\end{figure}

\begin{figure}[H]
\includegraphics[width=18cm,height=18cm,keepaspectratio]{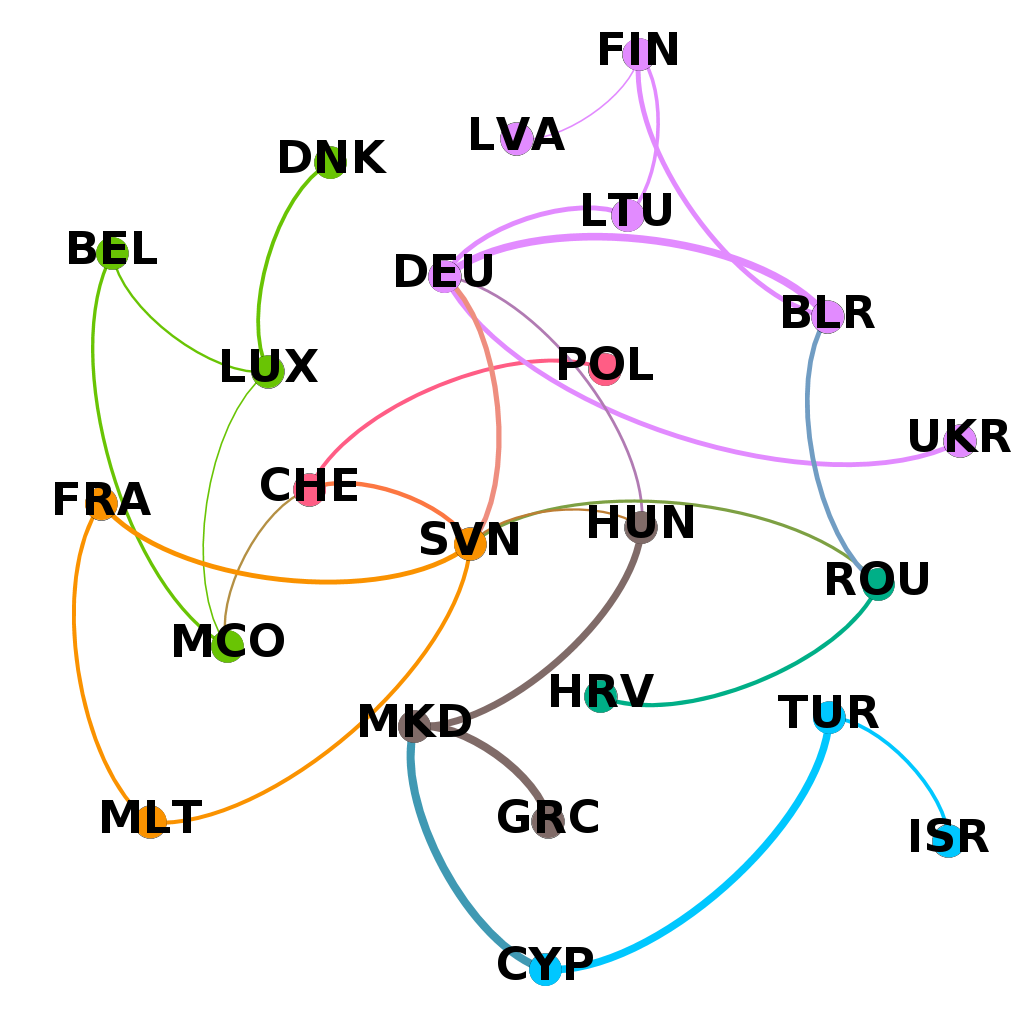}

\caption{Bidirectional neglect from the start of the competition.\label{fig:Bidirectional-neglect-from-1956}}
\end{figure}

\begin{figure}[H]
\includegraphics[width=1\paperwidth,height=0.4\textheight,keepaspectratio]{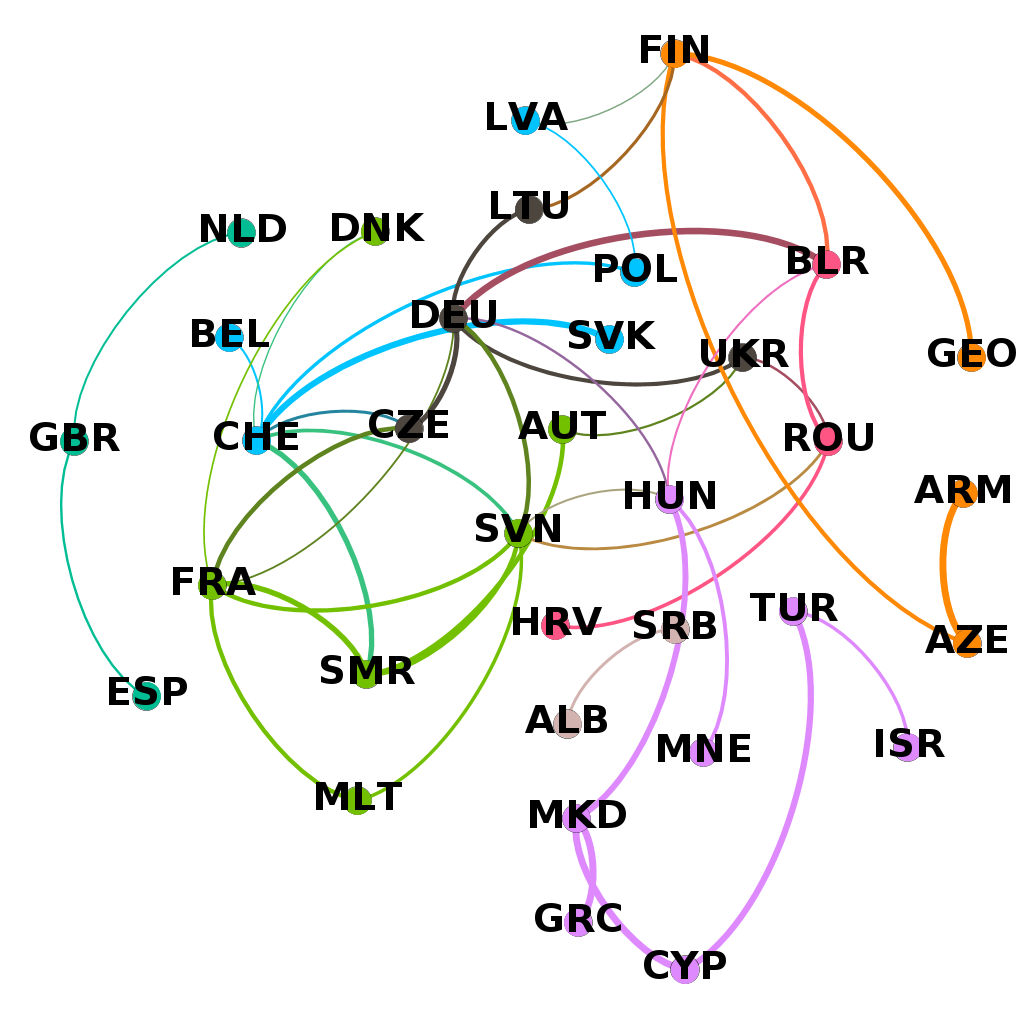}

\caption{Bidirectional neglect from the last 30 years (1989-2019).\label{fig:Bidirectional-neglect-30}}
\end{figure}

\begin{figure}[H]
\includegraphics[width=1\paperwidth,height=0.5\textheight,keepaspectratio]{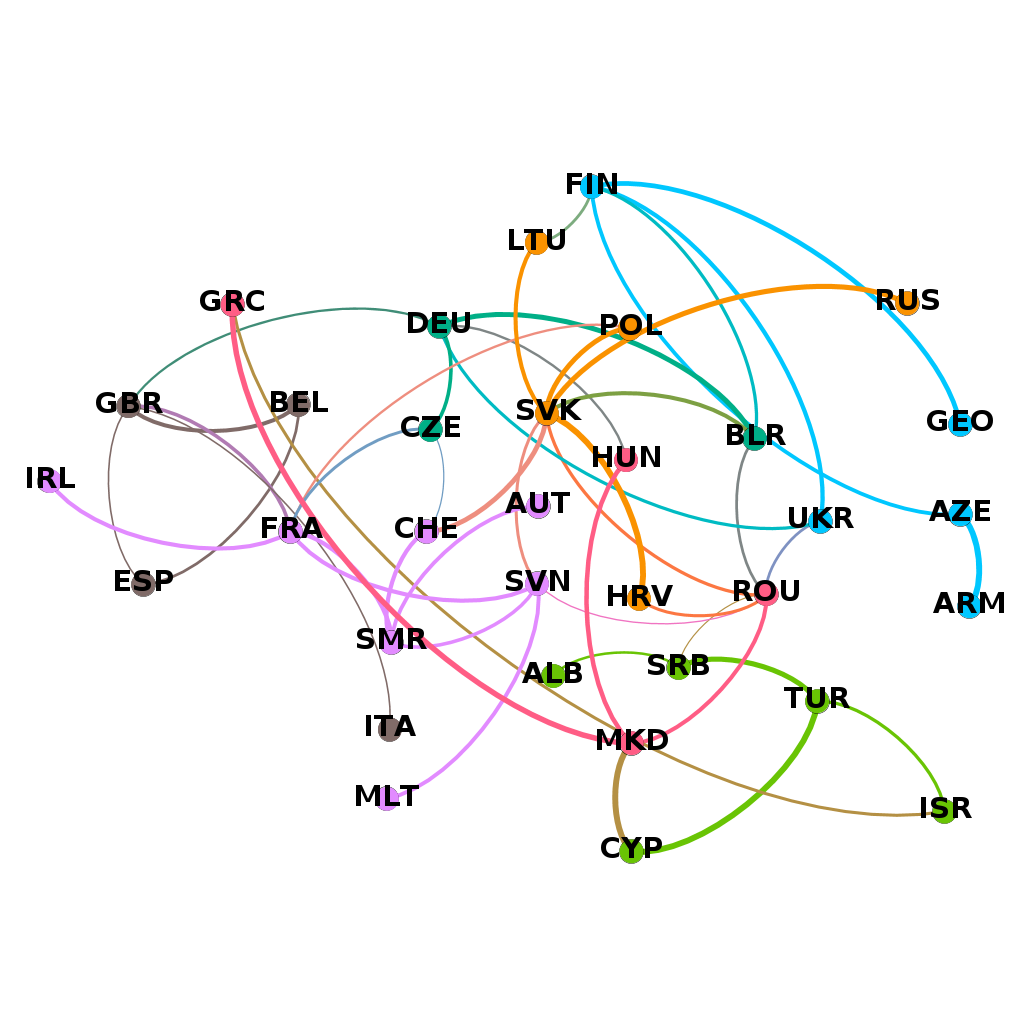}

\caption{Bidirectional neglect from the last 10 years (2009-2019).\label{fig:Bidirectional-neglect-10}}
\end{figure}

\section{Correlation between the community structure and success}

\begin{table}[H]
\begin{tabular}{|c|c|c|c|c|c|c|}
\hline 
 & \multicolumn{2}{c|}{bidirectional bias} & \multicolumn{2}{c|}{unidirectional bias} & \multicolumn{2}{c|}{bidirectional neglect}\tabularnewline
\hline 
Period  & Average  & Std. deviation  & Average  & Std. deviation  & Average  & Std. deviation\tabularnewline
\hline 
1  & 0.221899  & 0.100893  & 0.272000  & 0.183325  & 0.028571  & 0.045175\tabularnewline
5  & 0.182377  & 0.088343  & 0.213500  & 0.146715  & 0.008819  & 0.025185\tabularnewline
10  & 0.178759  & 0.091834  & 0.203822  & 0.149438  & 0.025190  & 0.040698\tabularnewline
15  & 0.146309  & 0.085600  & 0.183874  & 0.128531  & 0.030836  & 0.046235\tabularnewline
20  & 0.167683  & 0.092946  & 0.168643  & 0.126064  & 0.045053  & 0.063087\tabularnewline
25  & 0.160396  & 0.082057  & 0.176515  & 0.122498  & 0.061314  & 0.073130\tabularnewline
30  & 0.172900  & 0.087606  & 0.169401  & 0.114682  & 0.068828  & 0.077786\tabularnewline
35  & 0.154342  & 0.073746  & 0.152520  & 0.109934  & 0.079588  & 0.083365\tabularnewline
40  & 0.145658  & 0.068036  & 0.140207  & 0.097547  & 0.098842  & 0.100601\tabularnewline
45  & 0.144391  & 0.058237  & 0.139352  & 0.101466  & 0.119286  & 0.110119\tabularnewline
50  & 0.151423  & 0.062818  & 0.137625  & 0.094858  & 0.128911  & 0.112514\tabularnewline
60  & 0.137798  & 0.060983  & 0.172797  & 0.113359  & 0.100064  & 0.074564\tabularnewline
63  & 0.148520  & 0.063798  & 0.118017  & 0.062083  & 0.089192  & 0.060707\tabularnewline
\hline 
\end{tabular}\caption{Average percentage of points from cluster \label{tab:AVGpercentagePointsFromCluster}}
\end{table}

\begin{table}[H]
\begin{tabular}{|c|c|c|c|c|c|c|}
\hline 
 & \multicolumn{2}{c|}{bidirectional bias} & \multicolumn{2}{c|}{unidirectional bias} & \multicolumn{2}{c|}{bidirectional neglect}\tabularnewline
\hline 
Period  & Average  & Std. deviation  & Average  & Std. deviation  & Average  & Std. deviation\tabularnewline
\hline 
1  & 4.016861  & 8.845025  & 5.651971  & 10.576302  & 1.000000  & 0.000000\tabularnewline
5  & 11.277358  & 12.190952  & 11.026810  & 12.826616  & 11.243243  & 7.414031\tabularnewline
10  & 14.761141  & 11.743697  & 13.290155  & 12.419042  & 19.939394  & 10.590783\tabularnewline
15  & 15.379965  & 11.214536  & 13.944306  & 11.625325  & 22.351812  & 10.998832\tabularnewline
20  & 15.294872  & 10.934079  & 14.337316  & 11.244574  & 23.031690  & 11.220221\tabularnewline
25  & 15.568889  & 10.940346  & 14.130389  & 10.757622  & 23.863777  & 11.516176\tabularnewline
30  & 16.186901  & 10.755275  & 13.700739  & 10.276011  & 24.791541  & 11.303505\tabularnewline
35  & 16.421986  & 10.351069  & 13.672965  & 10.068184  & 25.268405  & 11.174442\tabularnewline
40  & 16.538934  & 10.029798  & 13.572711  & 9.798170  & 25.789303  & 11.201456\tabularnewline
45  & 16.911271  & 10.138811  & 13.414918  & 9.767238  & 26.289908  & 11.634037\tabularnewline
50  & 16.835694  & 9.870618  & 13.766871  & 9.888998  & 27.171171  & 12.027089\tabularnewline
60  & 17.774194  & 9.883873  & 16.621622  & 11.320120  & 28.240000  & 11.567126\tabularnewline
63  & 18.470588  & 10.452231  & 16.538462  & 11.767722  & 28.419355  & 11.555683\tabularnewline
\hline 
\end{tabular}\caption{Average place \label{tab:AVGPlace}}
\end{table}

\begin{table}[H]
\begin{tabular}{|c|c|c|c|c|c|c|}
\hline 
 & \multicolumn{2}{c|}{bidirectional bias} & \multicolumn{2}{c|}{unidirectional bias} & \multicolumn{2}{c|}{bidirectional neglect}\tabularnewline
\hline 
Period  & Average  & Std. deviation  & Average  & Std. deviation  & Average  & Std. deviation\tabularnewline
\hline 
1  & 192.798962  & 121.296850  & 145.539746  & 112.241151  & 10.000000  & 0.000000\tabularnewline
5  & 532.971698  & 272.120966  & 377.071046  & 243.509404  & 179.527027  & 122.623156\tabularnewline
10  & 812.572193  & 402.081197  & 639.287195  & 350.747929  & 334.397306  & 216.890460\tabularnewline
15  & 1016.191710  & 438.225082  & 876.906899  & 386.957619  & 508.735608  & 353.116841\tabularnewline
20  & 1160.397436  & 466.326237  & 1087.883272  & 421.469262  & 656.589789  & 458.733927\tabularnewline
25  & 1312.755556  & 493.326058  & 1320.077813  & 440.165090  & 800.803406  & 566.225431\tabularnewline
30  & 1479.075080  & 537.684058  & 1546.444581  & 473.229353  & 906.910876  & 632.688740\tabularnewline
35  & 1642.329787  & 586.564189  & 1754.513081  & 523.745983  & 997.437117  & 693.322978\tabularnewline
40  & 1816.186475  & 652.284413  & 1959.542190  & 586.185619  & 1087.307942  & 752.997564\tabularnewline
45  & 1964.357314  & 704.277567  & 2164.620047  & 651.111263  & 1228.269725  & 812.415801\tabularnewline
50  & 2120.864023  & 728.283705  & 2333.263804  & 711.741826  & 1358.369369  & 853.760668\tabularnewline
60  & 2631.717742  & 721.080927  & 2699.945946  & 875.262605  & 1884.672000  & 829.750447\tabularnewline
63  & 2877.676471  & 732.652824  & 2995.423077  & 925.767630  & 2151.290323  & 871.485243\tabularnewline
\hline 
\end{tabular}\caption{Average points \label{tab:AVGpoints}}
\end{table}

\begin{figure}[H]
\includegraphics[width=1\textwidth]{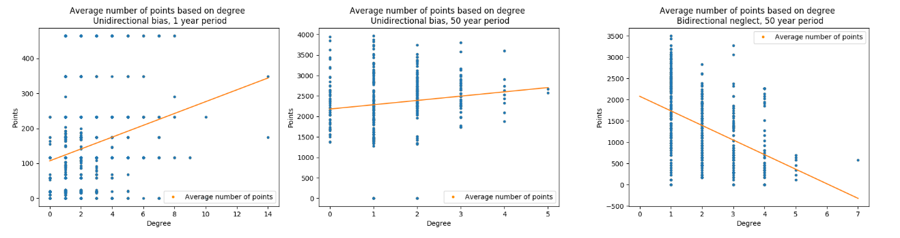}

\caption{Relationship between the degree in the bidirectional bias (left and
center) and neglect (right) and the number of points received in the
last 50 years.\label{fig:Degree-points-50}}
\end{figure}

\end{document}